\documentclass[notitlepage, showpacs, nobalancelastpage, twocolumn, aps,prb,superscriptaddress]{revtex4-2}
\usepackage[english]{babel}
\RequirePackage[T1]{fontenc}
\RequirePackage{times}
\usepackage{pifont}
\usepackage{siunitx}
\usepackage[italicdiff]{physics}
\usepackage{amsfonts}
\usepackage{subfigure}
\usepackage{amsmath}
\usepackage{amssymb}
\usepackage{graphicx, epstopdf}
\usepackage{appendix}
\usepackage{xspace}
\usepackage{lipsum}
\usepackage{array}
\usepackage{hyperref}
\usepackage[dvipsnames]{xcolor}
\usepackage{my_symbols}
\usepackage{physics}
\usepackage{CJK}
\usepackage{bm}
\usepackage{tikz}
\usepackage{booktabs}
\usepackage{multirow}
\usepackage{pifont}
\usepackage{setspace}

\begin{document}
\begin{CJK*}{UTF8}{gbsn} 

\title{Magnonic radar for dynamic domain walls in synthetic antiferromagnets}

\author{Yang Zhang (张扬)}
\author{Jin Lan (兰金)}	
\email[Corresponding author:~]{lanjin@tju.edu.cn}
\affiliation{Center for Joint Quantum Studies and Department of Physics, School of Science, Tianjin University, 92 Weijin Road, Tianjin 300072, China}
\affiliation{Tianjin Key Laboratory of Low Dimensional Materials Physics and Preparing Technology, Tianjin University, Tianjin 300354, China}

\begin{abstract}
Spin wave and magnetic domain wall are two of basic excitations in magnetic systems, and their spatiotemporal interplay encodes rich information of underlying magnetic interactions.
In synthetic antiferromagnets, the domain wall acquires an inertia and the spin wave  unlocks the full polarization degree of freedom, lays a salient platform for their interplay.
Here we show that both the translational and angular velocities of domain wall in synthetic antiferromagnets can be detected via the scattered spin wave, through the synergy of translational and angular Doppler effects. 
Following the setup of an electromagnetic or acoustic radar, the time evolution of a domain wall state are accessible via a series of spin wave packets, in both non-invasive and invasive fashion.
The inspections in frequency domain, offer new paradigms in exploration and exploitation of magnetic excitations.
\end{abstract}

\maketitle
\end{CJK*}

\section{Introduction}

Spin waves, the collective excitations of ordered magnetizations, 
extend the applicability of spintronics via the ability to propagate in both magnetic metals and insulators \cite{kajiwara_transmission_2010,cornelissen_longdistance_2015,lebrun_tunable_2018,flebus_2024_2024}. 
Their broad frequency spectrum and pronounced nonlinearity \cite{zheng_tutorial_2023a,wang_deeply_2023,an_emergent_2024} enable diverse frequency transformations including shifts \cite{vlaminck_currentinduced_2008}, up-/down-conversions \cite{leenders_canted_2024,zhang_terahertzfielddriven_2024,zhang_terahertz_2024, xu_frequency_2025,zhang_terahertz_2025}, multiplications \cite{koerner_frequency_2022,huang_extreme_2024}, and comb formations \cite{wang_magnonic_2021,xu_magnonic_2023,wang_enhancement_2024,lan_coherent_2025}. 
As the reciprocal-space counterpart to temporal dynamics, the frequency domain establishes a complementary and versatile framework to categorize and conquer magnetic phenomena across multiple timescales \cite{pirro_advances_2021}.

A fundamental frequency-domain phenomenon is the Doppler effect \cite{chen_microdoppler_2006},  occurring when spin waves interact with moving environments. For moving objects (antennas \cite{stancil_observation_2006},  domain walls \cite{zhang_propagating_2014,xia_doppler_2016}, fluxons \cite{dobrovolskiy_magnon_2019}, magnetic anisotropy boundaries \cite{hu_voltagecontrolled_2024}) or spin currents (mediated by electrons \cite{vlaminck_currentinduced_2008}, magnons \cite{yu_spinwave_2021}, or Dzyaloshinskii-Moriya interactions \cite{liu_electric_2011,yu_magnetic_2016,kikuchi_dzyaloshinskiimoriya_2016}), spin-wave frequency shifts manifest as asymmetric wavevectors between upstream and downstream propagation. 
Crucially, when non-reciprocal shifts exceed critical thresholds, spin waves exhibit event-horizon analogs (black/white holes) \cite{roldan-molina_magnonic_2017,doornenbal_spinwave_2019} and Cherenkov radiation \cite{yan_fast_2011,shekhter_vortex_2011}.

Due to inherent magnetic precession during propagation, spin waves simultaneously transport linear and angular momenta \cite{haldane_geometrical_1986,volovik_linear_1987,yu_polarizationselective_2018,yu_magnetic_2021}. 
Hence, beside the translational Doppler effect connected to the linear momentum, the spin wave is subject to the angular Doppler effect, as enforced by its ability to the change its angular momentum \cite{lavery_detection_2013,emile_rotational_2023}. 
However, investigation of spin-wave angular Doppler effects remains limited \cite{kim_propulsion_2014},  constrained primarily by: i) the lack of suitable interacting objects with rotational velocities comparable to translational speeds, and ii) unresolved challenges in decoupling both Doppler mechanisms during general translational-rotational motion.

\begin{figure*}[bt]
    \centering 
   \includegraphics[width=0.99\textwidth,trim= 55 20 35 15, clip]{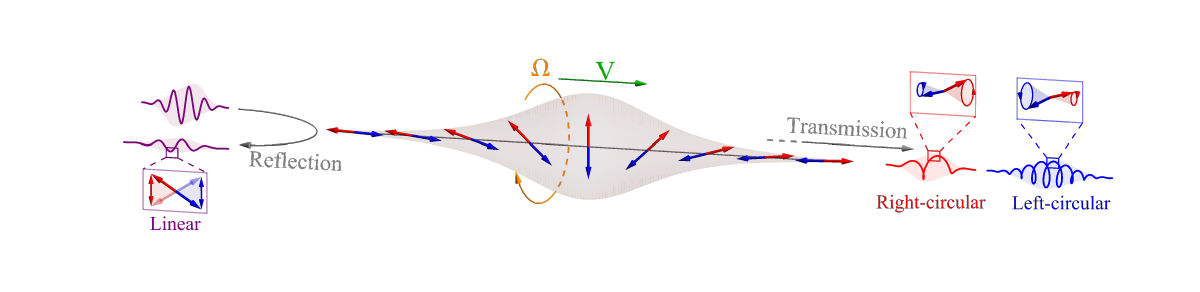}
    \caption{{\bf  Schematics of a magnonic radar.} 
   The red/blue arrows denote two sublattice magnetizations of an antiferromagnetic domain wall, and the green/orange arrows depict the direction of the translational/angular velocities.
   Through a simultaneously moving and precessing domain wall, a linearly-polarized spin wave packet splits into three wave packets with different frequencies and polarizations.     
    }
    \label{fig:radar}
\end{figure*}

In this work, we systematically investigate the Doppler frequency shifts of scattered spin waves induced by the dynamic state of a domain wall, and propose a magnonic radar scheme as illustrated in Fig. \ref{fig:radar}. 
A domain wall in a synthetic antiferromagnetic wire is adopted as the test target due to its capacity to sustain translational and rotational motion in low-dissipation environments, as well as its flexibility to interact with the spin wave.
Specifically, the reflected spin-wave frequency shift exclusively depends on the translational velocity of domain wall, while transmitted frequency shift depends on the both the translational and angular velocities as well as the spin wave polarization.
Using the spin wave spectra to probe the domain wall dynamics, establishes the frequency domain as a fundamental framework for all-magnetic information processing.

The remainder of this paper is organized as follows. 
Sec. II formulates the domain  wall dynamics in a synthetic antiferromagnetic wire, and establishes the connection between translation/angular velocities with domain wall tilt/offset between two magnetic sublayers. 
The translational and angular Doppler effects experienced by the scattered spin wave across the domain wall is theoretically formulated and numerically verified in Sec. III. 
Based on Doppler frequency shifts, magnonic radars established by sending and receiving successive spin wave packets  in both non-invasive and invasive fashion are inspected in Sec. IV. 
Discussions and conclusions are given in Sec. V.

\section{Domain wall Dynamics  in synthetic antiferromagnets}
\subsection{Domain wall profile}
 
Consider a synthetic antiferromagnetic wire extending along the $x$-direction, composed of two ferromagnetic sublayers coupled antiferromagnetically \cite{yang_domainwall_2015, duine_synthetic_2018}.
Denoting the magnetizations in the upper/lower layer as $\mb_{1/2}$, their magnetic dynamics are then governed by the coupled LLG equations \cite{lan_antiferromagnetic_2017} 
\begin{align}
    \label{eqn:LLG}
    \pdv{{\mb}_j}{t} = - \gamma \mb_j \times \bh_j + \alpha \mb_j \times \pdv{{\mb}_j}{t},
\end{align}
where $j = 1, 2$ labels the upper and lower sublayers. 
The effective magnetic field acting on the magnetization $\mb_j$ of each sublayer is given by 
\begin{equation}
    \bh_j = \frac{2}{\mu_{0}M_{s}} \left( A\nabla^{2}\mb_j + K m_j^z\hbz - J \mb_{\bar{j}} \right),
\end{equation}
where $\bar{j}$ denotes the complementary sublayer index (with $\bar{1} = 2$ and $\bar{2} = 1$). Here, $A$ represents the intralayer exchange stiffness, $J$ is the interlayer exchange coupling constant, and $K$ is the easy-axis anisotropy strength along the $\hbz$-direction.

Owing to the unity constraint $|\mb_j|=1$, the magnetization is expressed in spherical coordinates as $\mb_j \equiv (\sin\theta_j \cos\phi_j, \sin\theta_j \sin\phi_j, \cos\theta_j)$, where $\theta_j$ and $\phi_j$ denote the polar and azimuthal angles relative to the $\hbz$-axis. The ground state of the LLG equation \eqref{eqn:LLG} corresponds to uniform domain configurations with either $(\theta_1, \theta_2) = (0, \pi)$ or $(\theta_1, \theta_2) = (\pi, 0)$, resulting in $\mb_1 = -\mb_2 = \pm \hbz$.

The domain wall solution to the LLG equation \eqref{eqn:LLG} adopts a Walker-type profile characterized by \cite{yang_domainwall_2015}
\begin{align}
    \label{eqn:dm-theta}
    \theta_j = 2 \arctan \qty[\exp \qty(p_j \frac{x - X_j}{W})], \quad
    \phi_j = \Phi_j,
\end{align}
where $X_j$ and $\Phi_j$ represent the translational position and azimuthal angle of the domain wall in the $j$th sublayer, $p_j$ denotes the polarity ($p_1 = 1$ for upper and $p_2 = -1$ for lower sublayer), and $W=\sqrt{A/K}$ is the characteristic width.
At equilibrium, the sublayer magnetizations are antiparallel with $X_1 = X_2$ and $|\Phi_1 - \Phi_2| = \pi$.

\subsection{Domain wall dynamics}
Under slow evolution where domain walls maintain approximately fixed profiles, the parameters in Eq.~\eqref{eqn:dm-theta} are promoted  time-dependent collective coordinates \cite{lan_skew_2021}: $\mb_j(x, t) \equiv \mb_j[X_j(t), \Phi_j(t)]$. 
To better describe antiferromagnetic dynamics, we transform to symmetric and antisymmetric coordinates: The spatial coordinates become the  wall center  $X = (X_1 + X_2)/2$ and  position offset $\chi = X_1 - X_2$, while the angular coordinates transform to the  angular reference  $\Phi = (\Phi_1 + \Phi_2 - \pi)/2$ and angluar tilt  $\varphi = \Phi_1 - \Phi_2 + \pi$.

Following the standard derivation procedure for the Thiele equation \cite{thiele_steadystate_1973,tretiakov_dynamics_2008} and assuming small offset $\chi$ and tilt $\varphi$, the LLG equation \eqref{eqn:LLG} governing domain wall dynamics reduces to two separate groups:
(i) Coupled dynamics of the wall center $X$ and angular tilt $\varphi$, 
\begin{subequations}
    \label{eqn:dmw-X-varphi} 
    \begin{align}
        \label{eqn:dmw-X}
        \dot{X} - \alpha W \dot{\varphi} &= J W \varphi, \\
        \label{eqn:dmw-varphi}   
        \dot{\varphi} + \alpha \frac{\dot{X}}{W} &= 0,  
    \end{align} 
\end{subequations}
which exhibits ferromagnetic-like dynamics with $J$ acting as effective hard-axis anisotropy \cite{liu_spinjosephson_2016}. 
(ii) Coupled dynamics of the angular reference $\Phi$ and the position offset $\chi$, 
\begin{subequations}
    \label{eqn:dmw-Phi-chi} 
    \begin{align} 
        \label{eqn:dmw-Phi}
        \dot{\Phi} + \frac{\alpha}{W} \dot{\chi} &= -\frac{J\chi}{W}, \\
        \label{eqn:dmw-chi}
        \dot{\chi} - \alpha W \dot{\Phi} &= 0,
    \end{align}
\end{subequations}
which mirrors the structure of Eq.~\eqref{eqn:dmw-X-varphi} under the mapping $(\Phi, -\chi) \to (X, \varphi)$.

The steady solutions to the generalized Thiele equations \eqref{eqn:dmw-X-varphi} and \eqref{eqn:dmw-Phi-chi} are described by
\begin{subequations}
    \label{eqn:dmw_VOmega0}
    \begin{align}
        V &= \frac{J W \varphi_0}{1 + \alpha^2} e^{-t/  \tau}, \\
        \Omega &= -\frac{J \chi_0}{(1 + \alpha^2) W} e^{-t / \tau},
    \end{align}  
\end{subequations}
where the translational velocity $V \equiv \dot{X}$ and angular velocity $\Omega \equiv  \dot{\Phi}$ are governed by the initial offset $\varphi_0$ and tilt $\chi_0$, with $\tau = (1+\alpha^2)/\alpha J$ representing the relaxation time inversely proportional to the damping constant $\alpha$. 
For a  small magnetic damping $\alpha$, the domain wall maintains approximately constant translational and angular velocities $(V, \Omega)$ determined by the offset $\chi$ and tilt $\varphi$ throughout the long relaxation time $\tau$.

\begin{figure}[tb]
    \centering 
  \includegraphics[width=0.48\textwidth]{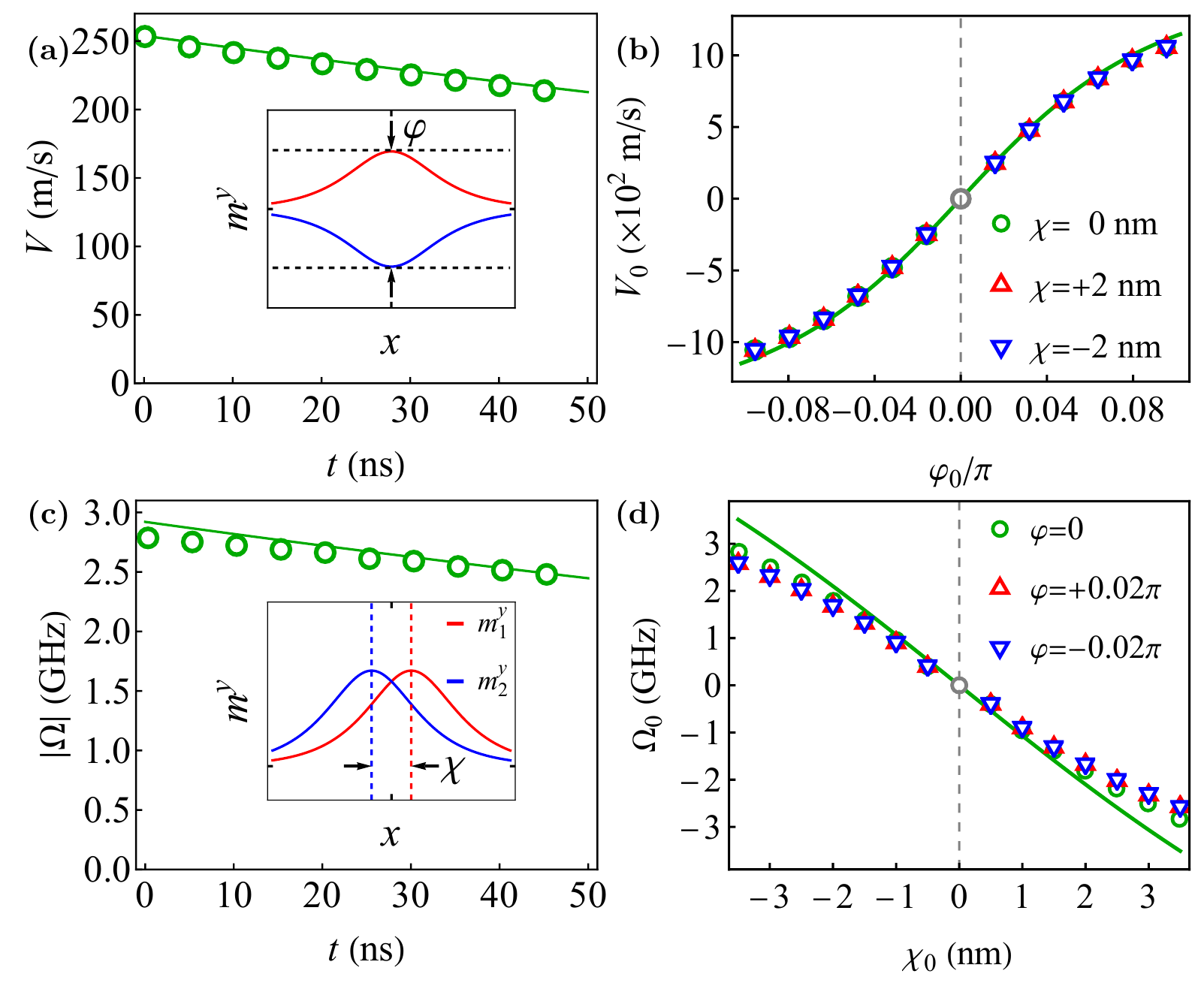}
    \caption{{\bf Domain wall velocities activated by an initial tilt $\varphi_0$ (a, b) or  offset $\chi_0$ (c, d).}
    (a) Time evolution of the translational velocity for tilt $\varphi_0=\SI{0.01}{\pi}$. 
    (b) Translational velocity as function of the initial tilt $\varphi_0$, for selectively chosen values of offset $\chi_0$. 
    (c) Time evolution of the angular velocity for offset $\chi_0=\SI{3}{nm}$. 
    (d) Angular velocity as function of the  initial offset $\chi_0$, for selectively chosen values of tilt $\varphi_0$. 
    In insets of figures (a, c), schematics of domain wall tilt and offset are illustrated, respectively.
    In all figures, the solid lines are based on the steady solution in \Eq{eqn:dmw_VOmega0}, and dots are extracted from micromagnetic simulations.
    }
    \label{fig:dmw_state} 
\end{figure}

The domain wall dynamics are further investigated through micromagnetic simulations using the open-source Mumax3 package \cite{vansteenkiste_design_2014}. The magnetic parameters employed in the simulations are:
The saturation magnetization $M_{s}=\SI{1.94e3}{A/m}$,  the intralayer exchange stiffness $A = \SI{4e-14}{J/m}$, the interlayer exchange constant $J=\SI{8e6}{J/m^3}$,  the easy-axis anisotropy $K = \SI{4e5}{J/m^{3}}$, and the damping constant $\alpha=\SI{1e-5}{}$. 
The synthetic antiferromagnetic wire has a length of \SI{2e4}{nm} with a lattice constant of \SI{10}{\nano\meter}, and domain wall is initialized at the wire center with $X = \SI{0}{nm}$.

For an initial tilt $\varphi_0 = \SI{0.01}{\pi}$, the domain wall acquires a translational velocity $V \approx \SI{250}{\meter/\second}$ in Fig.~\ref{fig:dmw_state}(a); while for an initial offset $\chi_0 = \SI{3}{\nano\meter}$, it develops an angular velocity $\Omega \approx \SI{3}{\giga\hertz}$ in Fig.~\ref{fig:dmw_state}(c). 
Both velocities exhibit identical relaxation behavior with a characteristic time $\tau \approx \SI{280}{\nano\second}$ for $\alpha = \SI{1e-5}{}$, consistent with the inverse proportionality to the damping constant. 
The dependence of translational velocity $V$ on tilt $\varphi_0$ and angular velocity $\Omega$ on offset $\chi_0$ is summarized in Fig.~\ref{fig:dmw_state}(b, d), showing excellent agreements with the steady-state solutions of Eq.~\eqref{eqn:dmw_VOmega0}.
The slight deviation from linear depedency on larger tilt $\varphi$ in Fig.~\ref{fig:dmw_state}(b) results from the Lorentz contraction of the antiferromagnetic domain wall, see Appendix \ref{sec:width_dw}. 
Moreover,  the magnitude and direction of translational velocity $V$ and angular velocity $\Omega$ can be flexibly and independently tuned across a wide range, via the selective control of the tilt $\varphi_0$ and offset $\chi_0$ parameters.

The dual-velocity tunability demonstrated in Fig.~\ref{fig:dmw_state}  establishes domain walls in synthetic antiferromagnetic wires as versatile platforms for investigating frequency-domain spin-wave scattering. 
The resulting scattering profiles, which exhibit unique spectral characteristics due to their coupling with the translational and rotational dynamics of the domain wall, will be systematically analyzed in subsequent sections.

\section{Spin wave scattering by dynamic domain walls}

\subsection{Spin wave dynamics} 

Due to the unity constraint, spin waves $\delta \mb_j$ are strictly transverse to the domain wall magnetization $\mb_j$, $\delta \mb_j \cdot \mb_j = 0$.
Adopting a local frame $\{\hbe_j^r, \hbe_j^\theta, \hbe_j^\phi\}$ attached to the domain wall with $\hbe_j^r \equiv \mb_j$, the spin wave is expressed as $\delta \mb_j = m_j^\theta \hbe_j^\theta + m_j^\phi \hbe_j^\phi$. 
In complex form,  the spin wave is alternatively written as $\psi_j^\sigma = m_j^\theta - i \sigma m_j^\phi$, where $\sigma = \pm 1$ denotes right- ($\sigma=+1$) or left-circular ($\sigma=-1$) polarization.
With above denotations,  the spin wave dynamics is recast from the LLG equation \eqref{eqn:LLG} to a Schr\"odinger-like equation \cite{lan_antiferromagnetic_2017}: 
\begin{align}
    \label{eqn:sw_eom}
    i \pdv{\psi_j^\sigma}{t} = \left[ -A \pdv[2]{}{x} + J + U(x) + i V \pdv{}{x} - \sigma m_j^z \Omega \right] \psi_j^\sigma 
    + J \psi_{\bar{j}}^\sigma,
\end{align}
where $U(x) = K \left[ 1 - 2 \sech^2 \left( x / W \right) \right]$ represents the effective potential generated by the inhomogeneous domain wall magnetization.

For the equilibrium configuration ($\chi = \varphi = 0$), the spin wave dynamics simplifies to \cite{lan_antiferromagnetic_2017}:    
\begin{align}
    \label{eqn:sw_eom_dw0}
    i \pdv{\psi_j^\sigma}{t}  = \left[-A \pdv[2]{}{x} + J + U(x) \right]\psi_j^\sigma 
    + J \psi_{\bar{j}}^\sigma,
\end{align}
where polarization dependence vanishes due to rotational symmetry. In the asymptotic region ($|x| \gg W$), the potential becomes uniform [$U(x) \to K$], yielding degenerate dispersion relations for both circular polarizations:
\begin{align}
    \omega = \gamma \sqrt{(Ak^2 + K + J)(Ak^2 + K)}
\end{align} 
with $\omega$ and $k$ denoting frequency and wavevector, respectively. 
The reflectionless nature of the P\"oschl-Teller potential $U(x)$ ensures perfect spin wave transmission through the domain wall for all frequencies and polarizations \cite{yu_polarizationselective_2018}.

When domain wall dynamics is activated through finite tilt ($\varphi \neq 0$) and offset ($\chi \neq 0$), the temporal evolution of the local frame $\{\hbe_j^r, \hbe_j^\theta, \hbe_j^\phi\}$ is incorporated in \Eq{eqn:sw_eom} via the operator substitution
\begin{align}
    \label{eqn:covariant_dt}
    \pdv{}{t} \rightarrow \pdv{}{t} - V \pdv{}{x} - i \sigma  p_j   m_j^z \Omega,
\end{align}
simultaneously encoding the domain wall's translational velocity ($V$) and angular velocity ($\Omega$). This dual-velocity coupling breaks translational and rotational symmetries, generating polarization-dependent scattering with significant reflection absent in the static case.

\subsection{Translational and angular Doppler effects}
 
An alternative and more convenient perspective, beyond the direct solution of the complex equation \eqref{eqn:sw_eom}, is afforded by transforming from the static laboratory frame to the dynamic domain wall frame:  
\begin{align}
    \psi_j^\sigma (x,t) \to \psi_j^\sigma \qty[x-X(t), t]e^{-i \sigma \Phi(t) m_j^z },
\end{align}
where spin waves co-process with the domain wall. For steady dynamics (with constant translational velocity $V = \dot{X}$ and angular velocity $\Omega = \dot{\Phi}$), the coordinate transformation to the domain wall frame is defined by
\begin{subequations}
    \label{eqn:dw_frame}
\begin{align}
    x' &= x - Vt, \label{eqn:coord_trans_x} \\
    t' &= \frac{\omega}{\omega - \sigma  p_j m_j^z  \Omega} t, \
    \label{eqn:coord_trans_t}
\end{align}
\end{subequations}
where $\omega$ denotes the laboratory-frame spin wave frequency. The spin wave frequency in the domain wall frame then becomes 
\begin{align}
    \label{eqn:omega_dmw_frame}
    \omega' &= \omega - V k - \sigma p_j  m_j^z \Omega,
\end{align}
revealing that spin waves experiences both translational and angular Doppler shifts. 
The polarization dependency in the angular shift reflects the intrinsic coupling between the spin wave polarization and the domain wall precession dynamics.

\subsection{Frequency shifts of scattered spin waves}  

After scattering, spin wave frequencies are restored by transforming from the domain wall frame back to the laboratory frame, via the connections established in \Eq{eqn:dw_frame}. 
Accounting for both forward and reverse frame transformations, the frequency shifts for transmitted and reflected components are given by \cite{hu_voltagecontrolled_2024}
\begin{subequations}
\label{eqn:D_omega}
\begin{align}
    \label{eqn:D_omega_t}
    \Delta\omega^\sigma_t &= \omega_t - \omega =2\sigma \Omega + V(k - k_t'), \\
    \label{eqn:D_omega_r}
    \Delta\omega_r &= \omega_r - \omega = -V (k + k_r'),
\end{align}
\end{subequations}
where $k$ denotes the incident wavevector (laboratory frame), and $k'_{r/t}$ represent the scattered wavevectors satisfying the domain wall frame dispersion relations: $\omega'_r = \omega - Vk$ for the reflected wave (on the left side) and $\omega_t' = \omega - Vk + 2\sigma\Omega$ for the transmitted wave (on the right side), as defined in Eq.~\eqref{eqn:omega_dmw_frame}. 
The frequency shift $\Delta \omega_r$ of the reflected spin wave depends solely on the domain wall velocity $V$, while $\Delta \omega^\sigma_t$ for the transmitted spin wave is additionally influenced by the angular velocity $\Omega$ through its coupling to the spin wave polarization.

\begin{figure}[tb]
\centering 
  \includegraphics[width=0.48\textwidth]{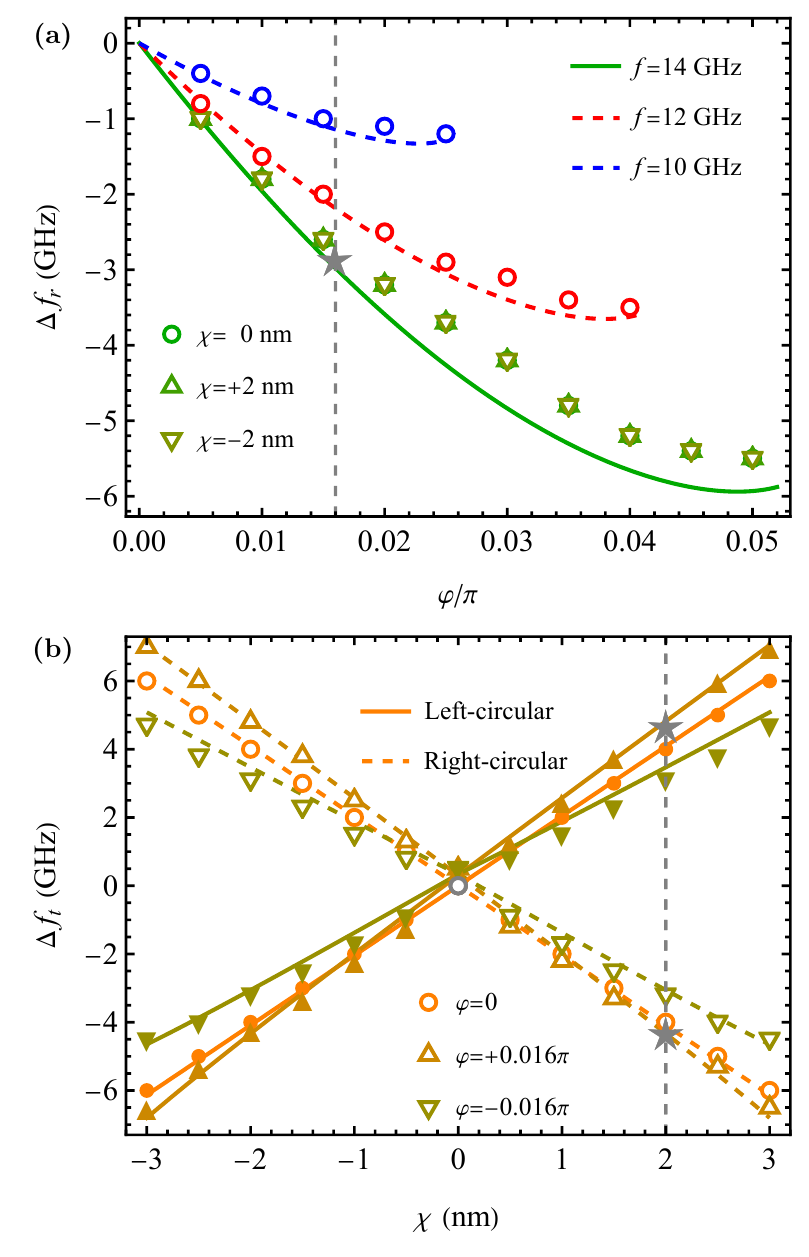}
  \caption{
  {\bf Frequency shifts of the reflected (a) and transmitted (b) spin wave as function of the domain wall tilt $\varphi$ and offset $\chi$.}
  The solid lines are based on theoretial equation \eqref{eqn:D_omega}, and the dots are extracted from micromagnetic simulations. 
  The spin wave is incident from the left side, and the  reflected/transmitted signals are monitored at selected points on the left-/right-side of the domain wall.
  The star with tilt $\varphi=\SI{0.016}{\pi}$ and offset $\chi=\SI{2}{nm}$ highlights the domain wall configuration adopted in Figs. \ref{fig:sw_packet_one} and \ref{fig:sw_packets}. }
  \label{fig:Doppler} 
\end{figure}

In micromagnetic simulations, the translational velocity $V$ and angular velocity $\Omega$ are controlled through the tilt $\varphi$ and offset $\chi$, respectively, following \Eq{eqn:dmw_VOmega0}. 
As shown in Fig.~\ref{fig:Doppler}(a), the simulated frequency shifts $\Delta f_r$ of reflected spin waves exhibit excellent agreement with theoretical equation \eqref{eqn:D_omega_r}, demonstrating two key characteristics: 
(i) linear scaling of shift magnitude with $\varphi$ (and thus $V$), and 
(ii) invariance under changes in $\chi$ (and consequently $\Omega$). 
Notably, only negative frequency shifts corresponding to forward-moving domain walls are observed, while positive shifts from backward motion remain inaccessible due to absence of spin wave reflection in that velocity regime.

Unlike the reflected component, transmitted spin waves exhibit bidirectional frequency shifts as transmission persists for both domain wall rotation directions, as demonstrated in Fig.~\ref{fig:Doppler}(b). 
The frequency shift $\Delta f_t$ shows three distinctive features: 
(i) approximate proportionality to displacement $\chi$ (controlling $\Omega$), 
(ii) additional modulation by tilt $\varphi$ (controlling $V$) with sign reversal for opposite polarizations, and 
(iii) asymmetry between positive and negative $\varphi$ values. 
This polarization-dependent shift originates from two collaborative mechanisms: 
a direct contribution from angular velocity $\Omega$ (via $\chi$), 
and an indirect contribution from wavevector shifts coupled to translational velocity $V$ (via $\varphi$). 
The observed $\varphi$-asymmetry arises because $V$ nonlinearly couples to the frequency shift through modifications of wavevector $k_t'$.

\section{Magnonic radar}

By leveraging frequency shifts in scattered spin-wave signals, a magnonic radar can be realized through active transmission and detection of single or multiple spin-wave packets, taking analogy to its electromagnetic or acoustic counterparts \cite{chen_microdoppler_2006}.

\subsection{Single spin wave packet}

\begin{figure}[tb]
    \centering 
    \includegraphics[width=0.48\textwidth,trim= 20 20 5 5, clip]{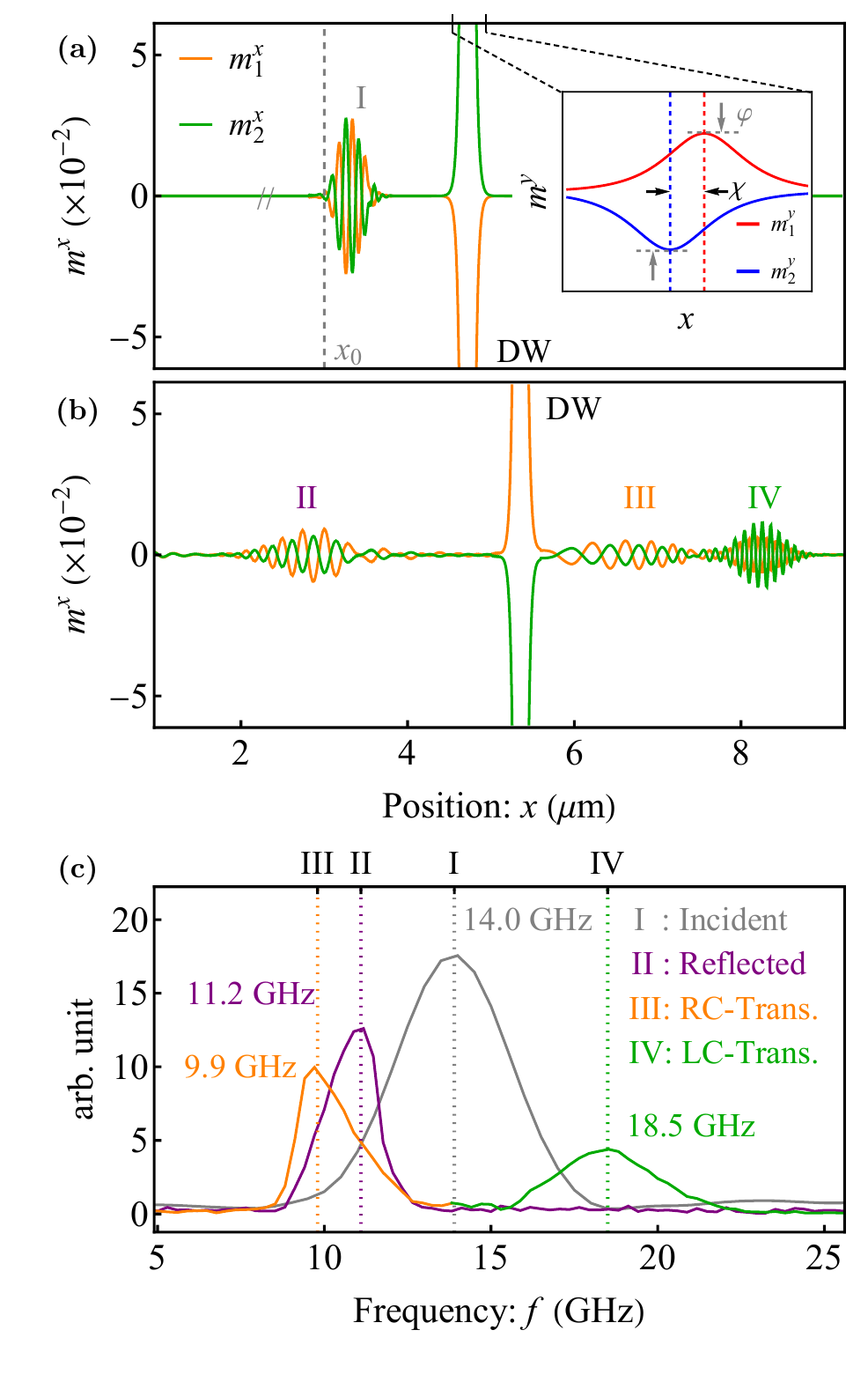}
    \caption{{\bf Probing dynamic states of a domain wall via a single spin wave packet.} 
    (a) Instantaneous profile of the incident spin wave packet and the domain wall. 
    Inset: schematic of the domain wall tilt and offset.
    (b) Instantaneous profile of the reflected and transmitted spin wave packets on two sides of the domain wall. 
    (c) Spectra of incident and scattered spin wave packets. 
    Gray and purple lines are for incident and reflected signals, and orange/green lines are for right-circular (RC) and left-circular (LC) components of the transmitted signals.
    }
    \label{fig:sw_packet_one} 
\end{figure}

To generate a spin wave packet, a time-dependent magnetic field is applied with $h_{\rm ext}(t)=h_0 \sin(2 \pi f_0 t) \exp[-(t-t_0)^2/2\zeta^2]$, where $h_0$ represents the amplitude of the excited wave packet, $f_0$ is the central frequency, $t_0$ is the central time, and $\zeta$ represents the half width. 
In Fig. \ref{fig:sw_packet_one}(a), an $x$-polarized spin-wave packet is prepared by an antenna at $x_0=\SI{3}{\um}$, through an oscillating magnetic field in Gaussian form with central frequency of $f_0=\SI{14}{GHz}$ and half width of $\zeta=\SI{1}{ns}$.
Simultaneously, Fig.~\ref{fig:sw_packet_one}(a) inset shows a dynamic domain wall configured with translational velocity $V=\SI{244}{m/s}$ and angular velocity $ \Omega=- \SI{1.76}{GHz}$ through specific tilt $\varphi=\SI{0.016}{\pi}$ and offset $\chi=\SI{2}{nm}$ (marked in Fig. \ref{fig:Doppler}). 
After scattering, the incident wave packet splits into three spatially separated packets in Fig. \ref{fig:sw_packet_one}(b): a reflected packet of linear polarization, and two transmitted packets of opposite circular polarizations.
Each packet exhibits distinct wavevectors and consequently propagates at different group velocities, enabling their spatial separation, similar to the scenario illustrated in Fig. \ref{fig:radar}.

The spectra of all three wave-packets, plotted collectively in Fig.~\ref{fig:sw_packet_one}(c), retain Gaussian profiles but exhibit distinct central frequencies.
The reflected spin wave shows a downward frequency shift of $\Delta f_r=\SI{-2.8}{GHz}$, while the transmitted left- and right-circularly polarized waves display shifts of $\Delta f_t^-=\SI{4.5}{GHz}$ and $\Delta f_t^+=\SI{-4.1}{GHz}$ in opposite directions (noting that $\Omega<0$). 
All these frequency shifts are consistent with the theoretical formulations in \Eq{eqn:D_omega}, as well as the specific domain wall configuration  in Fig. \ref{fig:Doppler}.  
Notably, the frequency shifts of transmitted signals are much larger than the angular velocity  in magnitude ($|\Delta f_t|>|\Omega|$), demonstrating the indispensable contribution of the translational Doppler effect in the transmitted signal.

\begin{figure*}[tb]
    \centering 
    \includegraphics[width=0.99\textwidth,trim= 20 5 2 5, clip]{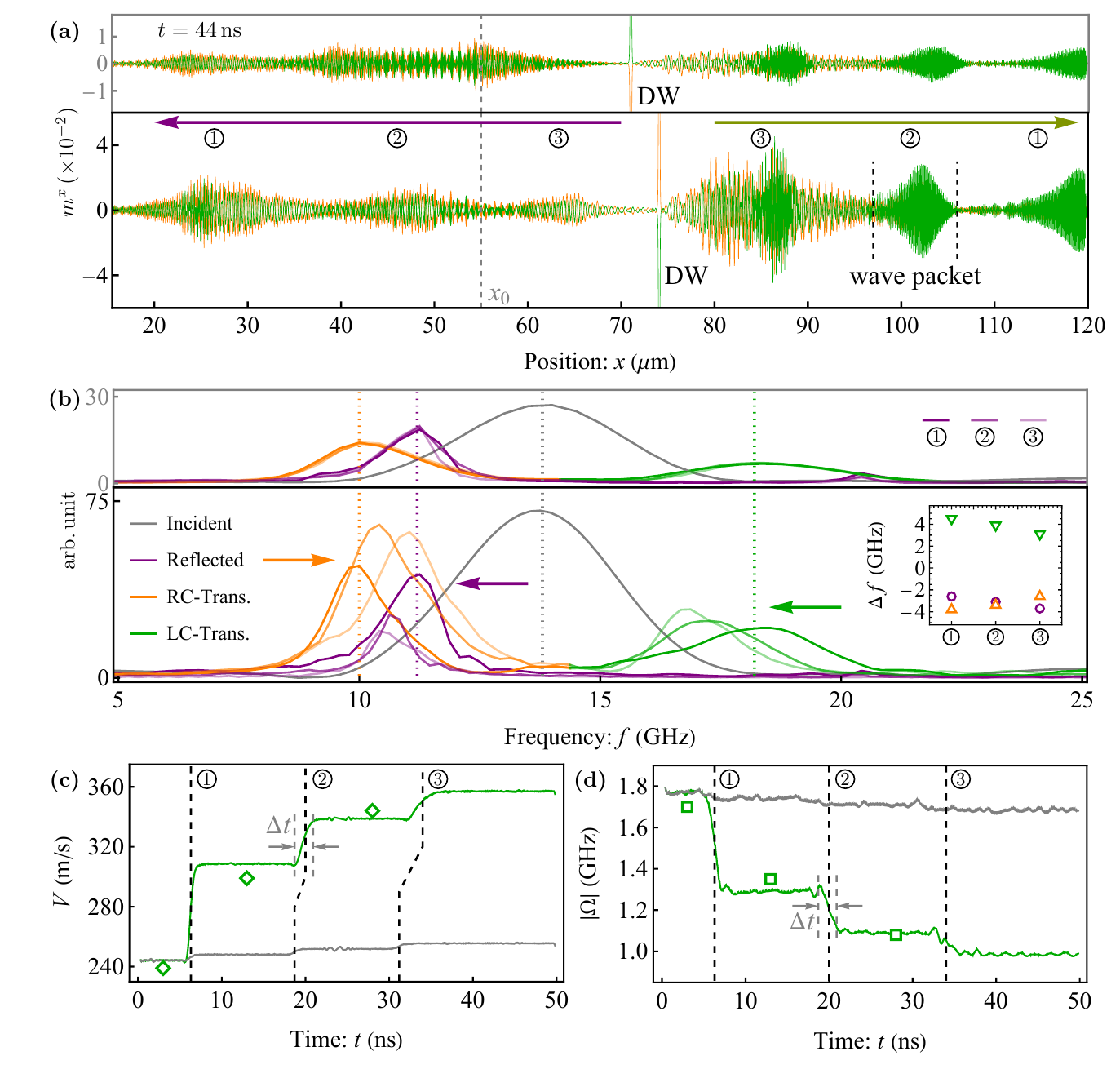}
    \caption{{\bf Probing dynamic states of a domain wall via series of spin wave packets.}
    (a) Instantaneous magnetization profile for three consequence spin wave packets scattered by a dynamic domain wall.
    Upper panel is for incident spin wave amplitude $0.04$ and lower panel is for amplitude $0.16$. 
    Both the reflected and transmitted spin wave packets are indexed by circled numbers. 
    (b) Spectra of the reflected and transmitted spin wave signals for small and large amplitude of the incident spin wave. 
    The gray/purple/orange solid lines are for spectra of the  incident/reflected/transmitted signals, for which the darker line corresponds to spin wave packets later in time.
    Inset: frequency shifts of successive spin wave packets. 
    (c, d) Time evolutions of the translational and angular velocities of domain wall. The solid lines are extraced from domain wall profile, and the dots are calculated from spin wave frequency shift. 
    The green/gray lines are for the case of spin wave with large/small amplitudes. 
    The vertical lines indexed with circled numbers indicate the time that the specific spin wave packet impinges on the domain wall.
   }
    \label{fig:sw_packets} 
\end{figure*}

\subsection{Multiple spin wave packets}

When extending from a single to three consecutive spin wave packets, Fig.~\ref{fig:sw_packets}(a) shows the spatial profiles of the instantaneous magnetization after scattering for both small and large amplitudes.
Due to the long propagation time ($\SI{44}{ns}$), all wave packets exhibit significant broadening, resulting in mixed circular polarization modes within the transmitted signals.
Fig.~\ref{fig:sw_packets}(b) displays the spectra of the three successive packets on the reflected and transmitted sides, revealing similar yet more intricate frequency shifts compared to the single-packet case in Fig.~\ref{fig:sw_packet_one}.
The temporal evolution of the domain wall's translational and angular velocities is further shown in Fig.~\ref{fig:sw_packets}(c, d), where the discrete values deduced from spin wave frequency shifts show good agreement with the velocities extracted continuously from the domain wall profile.

For small incident amplitude ($0.04$), successive packets maintain nearly identical magnetization profiles on both reflected and transmitted sides. 
Corresponding spectra show essentially perfect overlap with preserved central frequencies and peak amplitudes across successive packets. 
Stable scattering characteristics are corroborated by steady domain wall velocities, exhibiting only minor perturbations during packet passages.

\begin{figure*}[tb]
    \centering 
   \includegraphics[width=0.99\textwidth,trim= 10 5 35 5, clip]{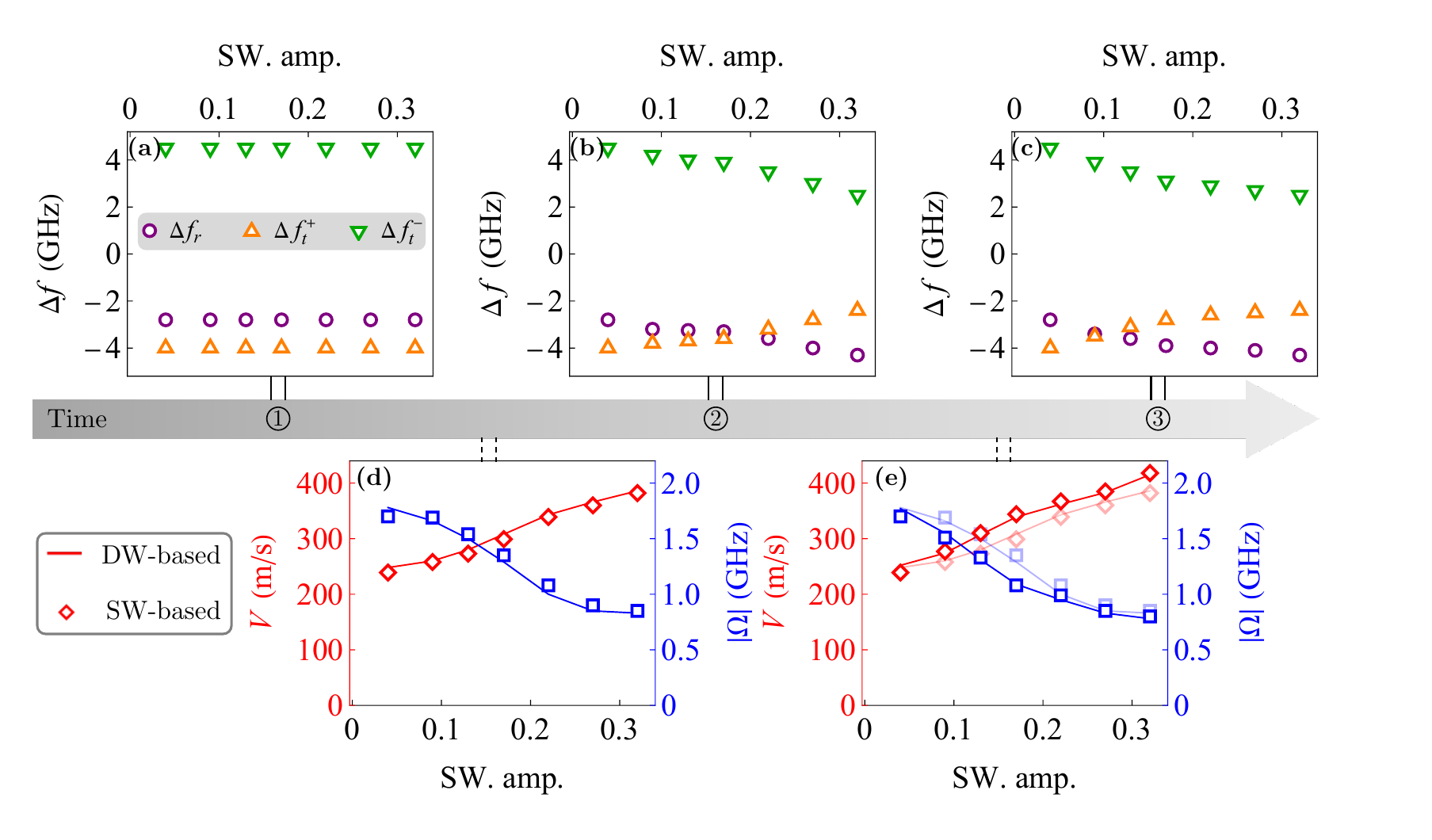}
    \caption{{\bf Spin wave frequency shifts (a-c) and domain wall velocities (d, e) as function of incident spin wave amplitudes, under successive penetration of  three spin wave packets. }
    In all figures, the dots are extracted from spin wave spectra, and the solid lines are analyzed from domain wall profiles.
    In (e), the shaded lines/dots are copied from (d) for easier comparisons. 
     }
    \label{fig:sw_amp} 
\end{figure*}

At elevated incident amplitude ($0.16$), both amplitude and width of reflected/transmitted packets progressively decline with successive passages, indicating dynamically evolving scattering. 
Reflected spectra exhibit successive downward central frequency shifts with gradual amplitude reduction, while transmitted spectra show progressive convergence toward incident frequency,  accompanied by narrowing peak width for right-circular polarization residing on high frequency side. 
These frequency shifts are summarized in Fig. \ref{fig:sw_packets}(b) inset, which coincide with pronounced domain wall velocity modifications during penetration: translational velocity increases from about $V\approx \SI{244}{m/s}$ to $V\approx \SI{357}{m/s}$, while angular velocity decreases from  $|\Omega|\approx \SI{1.76}{GHz}$ to $|\Omega|\approx \SI{0.99}{GHz}$.

Variations of spin wave scatterings are further evidenced by domain wall displacements in Fig.~\ref{fig:sw_packets}(a), showing approximately $\SI{5}{\mu m}$ displacement under large-amplitude excitation. 
Meanwhile, velocity transition duration increases from about $\SI{2}{ns}$ to $\SI{4}{ns}$ in Fig.~\ref{fig:sw_packets}(c, d), consistent with the higher domain wall speed. 
Notably, modification magnitude of domain wall velocity progressively diminishes,  due to increasing translational and rotational inertia of antiferromagnetic domain walls at elevated velocities \cite{shiino_antiferromagnetic_2016,yu_polarizationselective_2018, gomonay_Structure_2024}.

The contrasting responses in spin wave frequency and domain wall velocity for small ($0.04$) versus large ($0.16$) incident amplitudes demonstrate that the magnon radar scheme can operate in both non-invasive and invasive regimes.  
In the non-invasive configuration (small amplitude), spin wave signals remain steady throughout successive packets; while in the invasive configuration (large amplitude), cooperative evolution emerges between spin wave scattering and domain wall dynamics, enabling progressive spectral and kinematic modifications.

 Systematic variation of incident amplitude reveals distinct frequency shifts in successive spin wave packets, as summarized in Fig.~\ref{fig:sw_amp}. 
The first packet's frequency shift remains amplitude-independent [Fig.~\ref{fig:sw_amp}(a)], confirming the Doppler effect's intrinsic linearity. 
In contrast, the second packet exhibits significant frequency drift with increasing amplitude in Fig.~\ref{fig:sw_amp}(b), consistent with domain wall velocity changes following first-packet penetration in Fig.~\ref{fig:sw_amp}(d). 
While the third packet's frequency shift continues drifting, its magnitude diminishes in Fig.~\ref{fig:sw_amp}(c), due to reduced efficiency in modifying high-velocity domain wall states in Fig.~\ref{fig:sw_amp}(e).

\section{Discussions and Conclusions}

\emph{Hierarchy of magnonic Doppler effects.} 
As a magnetization structure, a domain wall hosts both global dynamics (translation/rotation) and local modes with internal nodes (e.g., breathing/twisting modes). 
The global dynamics generate the primary Doppler effects described by Eq.~\eqref{eqn:omega_dmw_frame}, while higher-order local dynamics produce micro-Doppler effects \cite{chen_microdoppler_2006}. These secondary effects exhibit significantly smaller magnitudes and emerge predominantly at large spin wave amplitudes. 
When domain wall velocities become time-dependent, the Doppler frequency shifts acquire corresponding temporal variations. 
This time dependence manifests as spectral redistribution \cite{schultheiss_time_2021,bar-hillel_time_2024,rao_timevarying_2025}, exemplified by frequency summation and differentiation under sinusoidal modulation. 
Collectively, these spatial and temporal dependencies establish complementary hierarchical dimensions for magnonic Doppler effects in dynamic domain walls.

\emph{Momentum implications of magnonic Doppler effects.}
Beyond amplitude variation \cite{kim_propulsion_2014,yu_polarizationselective_2018}, frequency shifts provide an additional indicator of spin-wave linear momentum changes. 
Pronounced frequency shifts in transmitted spin waves thus reveal momentum contributions beyond those from reflected waves. 
Furthermore, the transmitted frequency shift, primarily governed by domain wall angular velocity in Eq.~\eqref{eqn:D_omega_t}, manifests linear-to-angular momentum coupling through temporal breaking of rotational symmetry.

\begin{figure}[tb]
    \centering 
    \includegraphics[width=0.48\textwidth,trim= 10 5 20 20, clip]{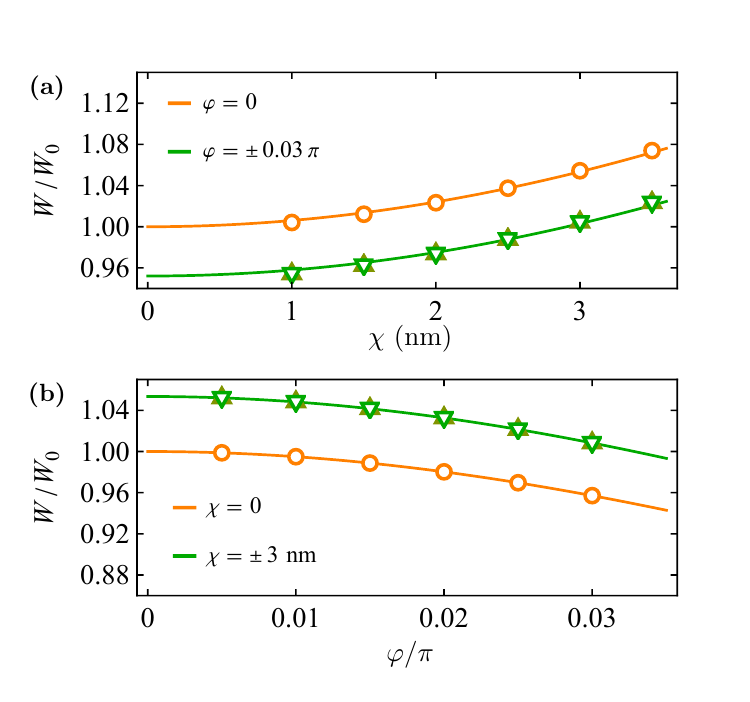}
    \caption{{\bf Normalized domain wall width $W/W_0$ as function of offset $\chi$ (a) and tilt $\varphi$ (b).} 
    In all figures, the dots are extracted from micromagnetic simulations, and the solid lines based on theoretical equation \eqref{eqn:limit-W}.
    }
    \label{fig:dw_W} 
\end{figure}

In conclusion, the translational and angular Doppler frequencies shifts of spin wave induced by a dynamic domain wall in a synthetic antiferromagnet wire are systematically investigated. 
By virtue of magnonic Doppler effects, a radar scheme that detects the domain wall state via transmitted and reflected spin waves are theoretically proposed and micromagnetically verified. 
The frequency degree of freedom, as a bridge between the slow and fast parts of magnetic excitations, opens new route toward full harnessing of magnetic systems.

\acknowledgements
This work is supported by National Natural Science Foundation of China (Grants No. 12374117 and No. 11904260) and Natural Science Foundation of Tianjin (Grant No. 20JCQNJC02020).

\appendix

\section{Velocity-dependent domain wall width}
\label{sec:width_dw}

Beside lowest order variations of domain wall in terms of  the translational positions $X_j$ and azimuthal angles $\Phi_j$  in  \Eq{eqn:dmw-X-varphi}, higher order variations in terms of width $W$ are also expected.  
According to the LLG equation \eqref{eqn:LLG}, the evolution of domain wall width is governed by 
\begin{align}
    \label{eqn:dynm-dm-W}
        \frac{\alpha \pi^2}{6 W} \dot{W} = \frac{AJ - V^2}{J W^2} - K + \frac{\Omega^2}{J},
\end{align}
where both translational and rotational dynamics are involved. 
The domain wall at steady state is given by  \cite{gomonay_Structure_2024} 
\begin{align}
    \label{eqn:limit-W}
    W =& \sqrt{\frac{1- \frac{V^2}{A J} }{1  - \frac{\Omega^2 W_0^2}{A J}}} W_0 = \sqrt{\frac{1- \frac{J \varphi^2}{A(4 + J \varphi^2)} }{1  - \frac{J \chi^2}{K(4 + J \chi^2)}}} W_0,
\end{align}
which is modulated by both translational velocity $V$ (or tilt $\varphi$) and angular velocity $\Omega$ (or offset $\chi$). 
As demonstrated in Fig. \ref{fig:dw_W},  the domain wall width under different parameter combinations ($\chi$, $\varphi$) extracted from micromagnetic simulations coincide exactly with the theoretical formulations in \Eq{eqn:limit-W}.
The domain wall contracts remarkably for elevated translational velocity $V$, while expands mildly for higher angular velocity $\Omega$.

\end{document}